\documentclass{nature}
\usepackage{caption}
\usepackage{amsmath,amssymb}
\usepackage{graphicx}

\title{Discovery of ubiquitous lithium production in low-mass stars}

\author{Yerra Bharat Kumar$^{1}$, Bacham E. Reddy$^{1,2}$ Simon W. Campbell$^3$, Sunayana Maben$^{1,4}$, Gang Zhao$^{1,4}$, Yuan-Sen Ting$^{5,6,7,8}$}

\begin{document}

\maketitle

\begin{affiliations}
 \item CAS Key Laboratory of Optical Astronomy, National Astronomical Observatories, Chinese Academy of Sciences, Beijing, China
 \item Indian Institute of Astrophysics, Department of Science \& Technology, Bangalore, India
 \item School of Physics and Astronomy, Monash University, Clayton, Victoria, Australia
 \item School of Astronomy and Space Science, University of Chinese Academy of Sciences, Beijing, China.
 \item Institute for Advanced Study, Princeton, NJ, USA
\item Department of Astrophysical Sciences, Princeton University, Princeton, NJ, USA
\item Observatories of the Carnegie Institution of Washington, Pasadena, CA, USA
\item Research School of Astronomy and Astrophysics, Australian National University, Canberra, Australian Capital Territory, Australia

\end{affiliations}





\begin{abstract}
The vast majority of stars with mass similar to the Sun are expected to only destroy lithium over the course of their lives, via low-temperature nuclear burning. This has now been supported by observations of hundreds of thousands of red giant stars\cite{brown1989,bharat2011,deepak2019,singh2019b,casey2019}. Here we perform the first large-scale systematic investigation into the Li content of stars in the red clump phase of evolution, which directly follows the red giant branch phase. Surprisingly we find that all red clump stars have high levels of lithium for their evolutionary stage. On average the lithium content increases by a factor of 40 after the end of the red giant branch stage. This suggests that all low-mass stars undergo a lithium production phase between the tip of the red giant branch and the red clump. We demonstrate that our finding is not predicted by stellar theory, revealing a stark tension between observations and models. We also show that the heavily studied\cite{brown1989,reddy2005,bharat2011,singh2019b,casey2019} very Li-rich giants, with A(Li)~$> +1.5$~dex, represent only the extreme tail of the lithium enhancement distribution, comprising 3\% of red clump stars. 
Our findings suggest a new definition limit for Li-richness in red clump stars, A(Li)~$> -0.9$~dex, which is much lower than the limit of A(Li)~$> +1.5$ dex used over many decades\cite{brown1989,castilho1995,reddy2005,carlberg2016,casey2019,holanda2020}.
\end{abstract}

The origin of giant stars highly enriched in lithium has been, and still is, a long-standing mystery\cite{wallerstein1982,brown1989,bharat2011,yan2018,casey2019}. Stellar lithium abundance is defined by A(Li), which is the logarithmic abundance of lithium: A(Li) = $\log_{10}(\rm{n(Li)/n(H)}) + 12$, where n(Li) and n(H) are the number densities of Li and H atoms. Stellar models predict lithium abundances in low mass ($< 2$~M$_{\odot}$) red giant branch (RGB) stars to be not more than A(Li)~$\sim$1.5~dex\cite{iben1967a}, due to depletion through the various evolutionary phases prior to this. Li-rich giants have therefore been defined as giant stars with A(Li)~$> 1.5$~dex. However, until recently, the exact evolutionary phase of Li-rich giants was unknown. Thanks to the advent of large-scale stellar surveys\cite{cui2012,buder2018} and asteroseismic observations\cite{bedding2011,vrard2016}, it has now become clear that the vast majority of Li-rich giants are not in the RGB phase, but in the stage of evolution directly after: the core-helium-burning `red clump' phase (RC)\cite{silvaaguirra2014,casey2019,deepak2019,singh2019b}. This new development raises a question: What value of A(Li) should be treated as normal among red clump giants? Or, in other words, what value of Li abundance qualifies RC giants to be categorised as Li-enriched?

To investigate this we compiled a database of 225,580 low-mass stars with quality surface temperatures from the Galactic Archaeology with HERMES (GALAH) survey's data release 2 (DR2)\cite{buder2018} and accurate luminosities from the Gaia survey's DR2\cite{GAIADR2Cat} (see Methods for more details). In Figure~1 we show this dataset, which was used to select our red clump (RC) sample (`RC box' in the figure). There is the possibility of contamination of this sample with red giant branch (RGB) stars since some RGB stars could overlap in the selected region of the HR diagram. To estimate the level of RGB star contamination requires knowledge of the evolutionary phases of the stars, through an independent method. To quantify the contamination we compared our RC box sample to a catalogue\cite{Ting2018} of inferred asteroseismic parameters that were derived using spectra from the LAMOST\cite{cui2012} and APOGEE\cite{albareti2017} surveys. We found 203 stars that overlap with our GALAH sample. The overlap sample is shown in the asteroseismic $\Delta P- \Delta\nu$ (period spacing - large frequency separation) diagram in Figure~2. Stars with $\Delta P \gtrsim 80$ seconds are considered RC stars\cite{bedding2011}, and the giants in the very narrow strip with $\Delta P \lesssim 80$ seconds are RGB stars\cite{vrard2016}. As shown in Figure~2,  the vast majority of the stars are in the RC phase, as they are well above the RGB limit of $\Delta P$ values. We confirm that the RGB contamination is low, with only $10\%$ of our GALAH RC box sample (Figure~1) found to be RGB stars. It is interesting to note that regardless of the degree of contamination it is clear from Figure~3 that practically all stars -- \textit{in all phases of evolution} -- have higher A(Li) than the RGB tip. This naturally includes all RC stars.

Figure~3 shows our sample from Figure~1 limited to stars with reliable lithium abundances (107,776 stars), in the lithium-luminosity plane. Overlaid is the RC sample (9,284 stars; 26,751 stars in total are giants, with $\log(g)<3.1$). Here the various lithium destruction phases can be seen. Observationally we can see that our sample of low-mass stars generally start at the main sequence turnoff with A(Li)~$\sim 2.5$~dex, then suffer depletion through the first dredge-up as they become red giants\cite{iben1967a}, moving from the left to right in this figure. This depletion is due to dilution as the envelope deepens and mixes up interior material heavily depleted in Li. After this the Li depletion rate slows, until the stars reach the RGB bump, at a luminosity of $\log_{10}(\rm{L/L_{\odot}})\sim 1.6$, where they start rapidly depleting Li again. This time the depletion is caused by Li destruction in the stellar envelope, through `deep mixing'. Deep mixing is a well-studied phenomenon whereby some proton-capture nucleosynthesis occurs just below the bottom of the giant star's convective envelope, destroying Li through the $^{7}\rm{Li}(p,\alpha)^{4}\rm{He}$ reaction. Theoretically it is modelled as thermohaline mixing\cite{charbonnel2007,Lattanzio2015}, a slow overturn of material that exposes Li to high enough temperatures to burn. This Li-destruction region connects to the convective envelope and hence the surface, where we observe the depletion. Deep mixing has been particularly well studied in globular clusters, which provide relatively homogeneous sample of stars useful for validating stellar models\cite{lind2009, Angelou2012,Dorazi2015,Kirby2016}. Previous work with globular clusters has led to this well established theoretical understanding of deep mixing\cite{charbonnel2007,Angelou2012,Lattanzio2015}.

The signature of deep mixing is clearly evident in our data, as Li drops from A(Li)~$\sim +0.5$~dex at the RGB bump to A(Li)~$\sim -0.5 \rightarrow -1.0$~dex at the RGB tip (Figure~3). The tip value is not well defined in our sample, since it does not contain the brightest stars, stopping about 0.5 magnitudes below the tip. After the RGB tip stars rapidly contract, evolving down in luminosity to the RC where they quiescently burn helium for about 100 million years. In the absence of a Li production mechanism it would be expected that the RC stars should show the same Li as the RGB tip, so the RC stars should lie directly below the RGB tip in the Li-luminosity diagram. To check this against theoretical expectations we calculated a set of solar-mass, solar-metallicity stellar models using the MESA code\cite{paxton2013}. Models with these parameters are representative since the mass of the giant stars in Figure~3 peaks at 1.0~M$_{\odot}$ ($\sigma = 0.4$~M$_{\odot}$), while the metallicity peaks at [Fe/H]~$=-0.1$ dex ($\sigma = 0.2$~dex). The models included thermohaline mixing on the RGB and were initialised to match the observed abundance of Li at the main sequence turn-off (A(Li)~$\sim 2.5$, Figure~3), after normal pre-main-sequence Li destruction. We found that a thermohaline mixing efficiency\cite{paxton2013} of $\alpha_{thm} = 100$ was required to match the late RGB Li depletion. Our $\alpha_{thm} = 100$ model reproduces the various observed phases of Li depletion well (dashed line in Figure~3). At the RGB tip the model has A(Li)~$=-0.9$, consistent with our observational data. As expected, the model then drops sharply in luminosity to the level of the RC, whilst maintaining the RGB tip A(Li). Shortly after, as the star evolves along the RC, Li is depleted further, with A(Li)~$= -1.3$~dex by the end of the RC evolution.

Turning our attention to the observed RC distribution, we find that it is peaked at A(Li)~$= +0.71$~dex ($\sigma = 0.39$). The vast majority of stars (within $2\sigma$) lie in the range of A(Li)~$= -0.07 \rightarrow 1.49$~dex. This is clearly very different to the model prediction for the RC, where the range is $-0.9 \rightarrow -1.3$~dex. The minimum difference between the model prediction (at the start of the RC) and the observed RC peak is 1.6~dex -- or a factor of 40 (Figure~3). This is a very large discrepancy between models and observations.

This observed RC Li distribution also highlights the rarity of the so-called `Li-rich giants', with A(Li)~$> 1.5$~dex, which have been the focus of the majority of the lithium in giants literature to date\cite{brown1989,reddy2005,bharat2011,yan2018,casey2019,singh2019b}. In our RC sample we find that $3.0\%$ of stars are above this limit (275 out of 9284 stars). In Figure~3 it can be seen that the commonly accepted value for defining Li-rich stars of A(Li)~$> 1.5$~dex is appropriate for lower RGB stars. However it is clearly not valid for more evolved stars, as has been noted before\cite{Kirby2016}, because of the strong Li depletion suffered by stars as they evolve up the RGB. The limit is particularly inappropriate for RC stars, which, as detailed above, should be taken as the RGB tip value of A(Li)~$> -0.9$~dex. This proposed limit for Li-richness in RC stars is 2.4~dex below that of the traditional limit which has been used for decades\cite{brown1989,bharat2011,casey2019}. We stress that this limit should be used in preference over the old one since it is now clear that Li-enhancement is primarily confined to the RC phase\cite{silvaaguirra2014,casey2019,singh2019b}.

 Observations of RC stars in open star clusters (OCs) lend support to our finding\cite{AnthonyTwarog2018,carlberg2016}. A recent study investigated lithium abundances in a sample of OCs at high spectral resolution\cite{carlberg2016}. Although OCs tend to host more massive stars, three of the studied OCs have stellar populations that can be considered low-mass, within our sample's mass range (M~$<2.0$~M$_{\odot}$). Remarkably, all three of these OCs were found to have average RC lithium abundances of A(Li)~$> +0.5$~dex.
 We include these observations in Figure~3 for comparison with our RC sample. Evolved RGB stars are rare in OCs, but there is some evidence that Li depletion along the RGB also occurs in  OCs\cite{pilachowski1986}. The recent high-resolution study, mentioned above, observed at most one evolved RGB star amongst their three OCs (this star could be either an RGB or AGB star). It was found to have a lithium abundance of A(Li)~$\lesssim -0.95$~dex. If this an evolved RGB star, then this would be quantitatively consistent with the finding for evolved RGB stars in our large sample (A(Li)~$\lesssim -0.9$~dex), and also with evolved RGB stars in globular clusters\cite{Kirby2016}. In summary, we speculate that a very similar level of Li enrichment is also occurring in open cluster stars between the RGB tip and RC phase. Clearly more observations of evolved RGB stars in OCs are needed.

The discovery reported here, that all low-mass RC stars are enhanced in lithium -- contrary to expectations from stellar theory -- suggests that some physical process is missing in the stellar models. It is clear from our data that a lithium production phase must occur at some time between the RGB tip and the RC phase. This places strict constraints on the possible mechanism. Critically, it appears to be a standard event, occurring in all low-mass stars. We note that our model in Figure~3 undergoes a slight Li production phase at the very top of the RGB. A(Li) only increases by 0.1~dex, but it may represent a clue as to a possible Li-enrichment mechanism and deserves further investigation. Finally, the existence of a tail of very Li-rich stars in the RC distribution, which are the traditional Li-rich giants (A(Li)~$> 1.5$~dex), suggests that the Li production event may be variable, or that there are other mechanisms\cite{aguileragomez2016,casey2019} for occasionally producing such extreme Li abundances.

\begin{methods}

\subsection{Sample Selection}

Assigning evolutionary phases to individual field stars unambiguously has been a long-standing problem. This is particularly difficult for the stars in the RC phase, since they lie close to the red giant branch in the  Hertzsprung-Russel diagram\cite{bharat2011,bedding2011}. One of the best ways to resolve this issue is to  have a sample of giants from star clusters as they have the same age and metallicity, and its members define stellar phases accurately in the $T_{\rm{eff}}$-luminosity (or color-magnitude) diagram. However, measurements of Li abundances of cluster stars systematically covering all the phases of low-mass giant evolution, such as the first dredge-up, luminosity bump, upper RGB, and the red clump, are currently lacking. So far studies of Li in globular clusters have focused primarily on the RGB, covering from the luminosity bump to the upper RGB\cite{Dorazi2015,Kirby2016}. In open clusters some RC populations have been investigated\cite{AnthonyTwarog2018,carlberg2016} (Figure~3 includes some open cluster results), however very few evolved RGB stars have been observed, largely due to their rarity. An alternative to this approach is to employ large stellar datasets of field stars, since the number of stars can be large enough to discern all the key evolutionary phases.

In this study we take advantage of the large abundance data set from the GALAH spectroscopic survey DR2\cite{buder2018}. The GALAH data is unusual in that it is a high-resolution survey, and thus has abundance measurements of higher precision than most of the other large surveys.  Since its release,  GALAH DR2 spectroscopic abundances have been subjected to a number of studies including Li abundances in RGB stars\cite{deepak2019,deepak2020}. The GALAH catalogue contains stellar parameters and elemental abundances, including Li, which has been corrected for departures from local thermodynamic equilibrium\cite{buder2018}, for a large number of stars based on spectra with a resolution of $\rm{R} \approx 28,000$. The catalog also includes flags indicating reliability of the measurement of each derived parameter. Reliability of stellar parameters are flagged as $\rm{flag}_{\rm{cannon}}= 0$ to 9, zero being the most reliable. We adopted only the data for which $\rm{flag}_{\rm{cannon}} = 0$. 
We further limited the sample to low-mass giants with stellar mass, $\rm{M} \leq 2.0$ M$_{\odot}$. Masses were estimated using the standard formula: 
\begin{displaymath}
    \log(\rm{M/M_{\odot}}) = \log(\rm{L}/\rm{L_{\odot}}) + 4\log(\rm{T_{eff}}_{\odot}/\rm{T_{eff}})
    + \rm{log}g - \rm{log}g_{\odot}
\end{displaymath}
where $T_{\rm{eff}_{\odot}} = 5777$ K and $\log g_{\odot} = 4.44$ are our adopted solar values. For each star we used the $T_{\rm{eff}}$ and $\log g$ values from the GALAH catalogue, whilst the luminosity values were taken from Gaia DR2 astrometry\cite{GAIADR2Cat}. The above criteria gave a sample of 225,580 low mass stars with reliable stellar parameters which we use to identify the red clump region (see Figure~1).

Measured Li abundances in Galah DR2 have a flag index ranging from 0 to 8, with zero being the most recommended, 1 for values derived with 2$\sigma$ accuracy, 2 for values derived from a (usually reliable) estimate from The Cannon  (a data-driven method for determining stellar parameters and abundances from spectroscopic data)\cite{buder2018}, and 3 for values having flags 1 and 2. We restricted our Li abundance data to flag values less than or equal to 3, with the additional constraint that the measurement errors be $\leq$~0.3~dex. 
With these criteria, we have a total sample of 107,776 low-mass stars with reliable stellar parameters and Li abundances (Figure~3). A subset of these are RC stars (9,284 stars).

\subsection{Detectability and Reliability of Galah Lithium Measurements:}
\hfill\break
\indent The 6707~\AA~Li line can become so weak it is not detectable. The point at which the line is not measurable depends on a range of factors:  T$_{\rm{eff}}$, surface gravity, signal-to-noise ratio, spectral resolution, and the Li content itself. We have investigated possible bias in our RC sample that may have arisen through our process of excluding stars with non-optimal measurements (see Sample Selection above).

\noindent The Galah DR2 catalog provides Li abundances for all stars. It also provides flags for abundance measurement quality. As detailed above, our final sample is based on higher-quality data from Galah DR2. Specifically we limited the sample based on (i) quality stellar parameters (flag$_{cannon} = 0$), and (ii) quality Li measurements (flag$_{Li} \leq 3$ and err$_{Li} < 0.3$~dex). The first step should not bias the lithium distribution, and it restricts the sample to reasonable data, however the second step could have excluded many upper limit measurements (limits are not explicitly flagged in Galah DR2). First, to determine what proportion of stars was excluded we removed the lithium quality constraints. In Extended Fig.~1 we show this raw sample (50,690 giant stars) along with the sample used in Figure~3 (26,751 giants). From these numbers it can be seen that a large proportion, $\sim 47\%$, of giants were excluded due to non-optimal A(Li) measurements.

\noindent In Extended Fig.~1 we also include the Li versus T$_{\rm{eff}}$ detection limits. These limit curves were computed by considering the uncertainty in equivalent width (EW) measurements using a theoretical relation\cite{cayrel1988}, adopting Galah SNR ($50 - 100$) and spectral resolution ($R \sim 28,000$). We estimated a detection limit of EW = $3-5$~m\AA, changing with SNR. We then synthesised a grid of spectra, using the code $MOOG$\cite{sneden1973}, for a range of Li abundances (to vary line strength), a range of T$_{\rm{eff}}$ ($4000-5200$~K), and a range of gravity ($1.7-3.1$~dex). The limit curves reveal (Extended Figures~1~and~2) that only a tiny proportion of the excluded stars had A(Li) below the (nominal) detection threshold ($< 2\%$ at 5~m\AA; we note that some spectra have much higher SNR than considered in our test). The proportion is slightly lower for the RC sample ($\sim 1\%$). We also show the limits compared to the distribution of A(Li) at different stages of sub-sampling (Extended Fig.~2). This figure clearly shows again that we are not missing a significant number of very Li-poor stars due to upper limits (the stars are rejected for other reasons). It also shows that our final high-quality sample is representative of the larger sample. This reinforces our result that the RC Li distribution is indeed peaked well above the RGB tip value. Crucially the RC peak we find, at $+0.7$~dex, is much greater than the Galah RC detectability limit.

\noindent We made further checks to test the Galah Li abundances, using independent data from the literature. Firstly, as shown in Figure~3 and Extended Fig.~3, observations of RC stars in open clusters show a very similar high peak in A(Li). Although many of the OC stars have upper limit measurements, $40 - 50\%$ of them are confirmed measurements. This shows that there is a large proportion of open cluster RC stars that have relatively high Li. In a second test we cross-matched the entire Galah DR2 catalogue with the high-resolution Gaia-ESO survey\cite{Gilmore2012} (their DR3). We found Galah Li values to be very reliable, with an average offset of only $+0.06$~dex ($\sigma=0.37$; see Extended Fig.~4). Amongst the overlap sample we found two RC stars. In addition we cross-matched Galah DR2 with the LAMOST survey\cite{cui2012}. Here we also found two RC stars. LAMOST does not provide Li abundances so we derived our own Li abundances\cite{singh2019b,kumar2018}. We find these stars to both be very Li-rich due to the fact that LAMOST is a low-resolution survey, so only very Li-rich stars can be detected\cite{casey2019}.  Although there are only 4 RC stars in these cross-matches, they also all show excellent agreement with Galah (Extended Fig.~4).
Finally, as also shown in Figure~3, we made a comparison with a sample of RC stars with evolutionary state confirmed asteroseismically using data from the Kepler space telescope\cite{mosser2012}. This sample has Li abundances based on very high resolution spectra\cite{Takeda2017}.
Limiting the Kepler sample to our [Fe/H] range we found 28 RC stars for comparison (displayed individually in Extended Figure~3). This literature sample was also peaked well above their detectability limit, and above the RGB tip value that we find. Crucially only one out of the 28 measurements is an upper limit in this case, so this shows the true distribution of their sample, which is quite broad (error bar in Figure~3; see also Extended Fig.~3). We note that this sample was drawn from a single Kepler pointing (the original Kepler mission), so it is a biased stellar sample.

In summary, all the evidence above strongly suggests that the public data we have used for Li abundances, the Galah data release two\cite{buder2018} (DR2), are reliable.

\subsection{Stellar Model Initial Lithium Abundance:}
\hfill\break
\indent We chose the lithium abundance of our stellar model to match the peak of the Galah Li distribution at the main sequence turn-off (MSTO). Galah shows a wide range of A(Li) at the MSTO (Figure~3). This wide range is typical of field populations and the Galah sample is consistent with previous studies\cite{fulbright00,lambert91}. We note that our model in Figure~3 included pre-main-sequence (PMS) destruction. This destruction amounted to $\sim 0.2$~dex, so the initial pre-main-sequence model abundance used was A(Li)~$=2.80$~dex.

\end{methods}


\subsection{References}

\begin{thebibliography}{10}
\expandafter\ifx\csname url\endcsname\relax
  \def\url#1{\texttt{#1}}\fi
\expandafter\ifx\csname urlprefix\endcsname\relax\def\urlprefix{URL }\fi
\providecommand{\bibinfo}[2]{#2}
\providecommand{\eprint}[2][]{\url{#2}}

\bibitem{brown1989}
\bibinfo{author}{{Brown}, J.~A.}, \bibinfo{author}{{Sneden}, C.},
  \bibinfo{author}{{Lambert}, D.~L.} \& \bibinfo{author}{{Dutchover}, E.~J.}
\newblock \bibinfo{title}{{A search for lithium-rich giant stars}}.
\newblock \emph{\bibinfo{journal}{Astrophys. J. Suppl.}}
  \textbf{\bibinfo{volume}{71}}, \bibinfo{pages}{293--322}
  (\bibinfo{year}{1989}).

\bibitem{bharat2011}
\bibinfo{author}{{Kumar}, Y.~B.}, \bibinfo{author}{{Reddy}, B.~E.} \&
  \bibinfo{author}{{Lambert}, D.~L.}
\newblock \bibinfo{title}{{Origin of Lithium Enrichment in K Giants}}.
\newblock \emph{\bibinfo{journal}{Astrophys. J.}}
  \textbf{\bibinfo{volume}{730}}, \bibinfo{pages}{L12} (\bibinfo{year}{2011}).

\bibitem{deepak2019}
\bibinfo{author}{{Deepak}} \& \bibinfo{author}{{Reddy}, B.~E.}
\newblock \bibinfo{title}{{Study of Lithium-rich giants with the GALAH
  spectroscopic survey}}.
\newblock \emph{\bibinfo{journal}{Mon. Not. R. Astron. Soc.}}
  \textbf{\bibinfo{volume}{484}}, \bibinfo{pages}{2000--2008}
  (\bibinfo{year}{2019}).

\bibitem{singh2019b}
\bibinfo{author}{{Singh}, R.}, \bibinfo{author}{{Reddy}, B.~E.},
  \bibinfo{author}{{Bharat Kumar}, Y.} \& \bibinfo{author}{{Antia}, H.~M.}
\newblock \bibinfo{title}{{Survey of Li-rich Giants among Kepler and LAMOST
  Fields: Determination of Li-rich Giants{\textquoteright} Evolutionary
  Phase}}.
\newblock \emph{\bibinfo{journal}{Astrophys. J.}}
  \textbf{\bibinfo{volume}{878}}, \bibinfo{pages}{L21} (\bibinfo{year}{2019}).

\bibitem{casey2019}
\bibinfo{author}{{Casey}, A.~R.} \emph{et~al.}
\newblock \bibinfo{title}{{Tidal Interactions between Binary Stars Can Drive
  Lithium Production in Low-mass Red Giants}}.
\newblock \emph{\bibinfo{journal}{Astrophys. J.}}
  \textbf{\bibinfo{volume}{880}}, \bibinfo{pages}{125} (\bibinfo{year}{2019}).

\bibitem{reddy2005}
\bibinfo{author}{{Reddy}, B.~E.} \& \bibinfo{author}{{Lambert}, D.~L.}
\newblock \bibinfo{title}{{Three Li-rich K Giants: IRAS 12327-6523, 13539-4153,
  and 17596-3952}}.
\newblock \emph{\bibinfo{journal}{Astron. J.}} \textbf{\bibinfo{volume}{129}},
  \bibinfo{pages}{2831--2835} (\bibinfo{year}{2005}).

\bibitem{castilho1995}
\bibinfo{author}{{Castilho}, B.~V.}, \bibinfo{author}{{Barbuy}, B.} \&
  \bibinfo{author}{{Gregorio-Hetem}, J.}
\newblock \bibinfo{title}{{Analysis of the moderately Li-rich giant HD
  146850.}}
\newblock \emph{\bibinfo{journal}{Astron. \& Astrophys.}}
  \textbf{\bibinfo{volume}{297}}, \bibinfo{pages}{503} (\bibinfo{year}{1995}).

\bibitem{carlberg2016}
\bibinfo{author}{{Carlberg}, J.~K.}, \bibinfo{author}{{Cunha}, K.} \&
  \bibinfo{author}{{Smith}, V.~V.}
\newblock \bibinfo{title}{{Lithium Inventory of 2 M $_{\odot}$ Red Clump Stars
  in Open Clusters: A Test of the Helium Flash Mechanism}}.
\newblock \emph{\bibinfo{journal}{Astrophys. J.}}
  \textbf{\bibinfo{volume}{827}}, \bibinfo{pages}{129} (\bibinfo{year}{2016}).

\bibitem{holanda2020}
\bibinfo{author}{{Holanda}, N.}, \bibinfo{author}{{Drake}, N.~A.} \&
  \bibinfo{author}{{Pereira}, C.~B.}
\newblock \bibinfo{title}{{HD 150382: A Lithium-rich Star at the Early-AGB
  Stage?}}
\newblock \emph{\bibinfo{journal}{Astron. J.}} \textbf{\bibinfo{volume}{159}},
  \bibinfo{pages}{9} (\bibinfo{year}{2020}).

\bibitem{wallerstein1982}
\bibinfo{author}{{Wallerstein}, G.} \& \bibinfo{author}{{Sneden}, C.}
\newblock \bibinfo{title}{{A K giant with an unusually high abundance of
  lithium : HD 112127.}}
\newblock \emph{\bibinfo{journal}{Astrophys. J.}}
  \textbf{\bibinfo{volume}{255}}, \bibinfo{pages}{577--584}
  (\bibinfo{year}{1982}).

\bibitem{yan2018}
\bibinfo{author}{{Yan}, H.-L.} \emph{et~al.}
\newblock \bibinfo{title}{{The nature of the lithium enrichment in the most
  Li-rich giant star}}.
\newblock \emph{\bibinfo{journal}{Nature Astronomy}}
  \textbf{\bibinfo{volume}{2}}, \bibinfo{pages}{790--795}
  (\bibinfo{year}{2018}).

\bibitem{iben1967a}
\bibinfo{author}{{Iben}, I.~J.}
\newblock \bibinfo{title}{{Stellar Evolution.VI. Evolution from the Main
  Sequence to the Red-Giant Branch for Stars of Mass 1 M\_$\{$sun$\}$, 1.25
  M\_$\{$sun$\}$, and 1.5 M\_$\{$sun$\}$}}.
\newblock \emph{\bibinfo{journal}{Astrophys. J.}}
  \textbf{\bibinfo{volume}{147}}, \bibinfo{pages}{624--+}
  (\bibinfo{year}{1967}).

\bibitem{cui2012}
\bibinfo{author}{{Cui}, X.-Q.} \emph{et~al.}
\newblock \bibinfo{title}{{The Large Sky Area Multi-Object Fiber Spectroscopic
  Telescope (LAMOST)}}.
\newblock \emph{\bibinfo{journal}{Res. Astron. Astrophys.}}
  \textbf{\bibinfo{volume}{12}}, \bibinfo{pages}{1197--1242}
  (\bibinfo{year}{2012}).

\bibitem{buder2018}
\bibinfo{author}{{Buder}, S.} \emph{et~al.}
\newblock \bibinfo{title}{{The GALAH Survey: second data release}}.
\newblock \emph{\bibinfo{journal}{Mon. Not. R. Astron. Soc.}}
  \textbf{\bibinfo{volume}{478}}, \bibinfo{pages}{4513--4552}
  (\bibinfo{year}{2018}).

\bibitem{bedding2011}
\bibinfo{author}{{Bedding}, T.~R.} \emph{et~al.}
\newblock \bibinfo{title}{{Gravity modes as a way to distinguish between
  hydrogen- and helium-burning red giant stars}}.
\newblock \emph{\bibinfo{journal}{Nature}} \textbf{\bibinfo{volume}{471}},
  \bibinfo{pages}{608--611} (\bibinfo{year}{2011}).

\bibitem{vrard2016}
\bibinfo{author}{{Vrard}, M.}, \bibinfo{author}{{Mosser}, B.} \&
  \bibinfo{author}{{Samadi}, R.}
\newblock \bibinfo{title}{{Period spacings in red giants. II. Automated
  measurement}}.
\newblock \emph{\bibinfo{journal}{Astron. \& Astrophys.}}
  \textbf{\bibinfo{volume}{588}}, \bibinfo{pages}{A87} (\bibinfo{year}{2016}).

\bibitem{silvaaguirra2014}
\bibinfo{author}{{Silva Aguirre}, V.} \emph{et~al.}
\newblock \bibinfo{title}{{Old Puzzle, New Insights: A Lithium-rich Giant
  Quietly Burning Helium in Its Core}}.
\newblock \emph{\bibinfo{journal}{Astrophys. J.}}
  \textbf{\bibinfo{volume}{784}}, \bibinfo{pages}{L16} (\bibinfo{year}{2014}).

\bibitem{GAIADR2Cat}
\bibinfo{author}{{Gaia Collaboration}}.
\newblock \bibinfo{title}{{VizieR Online Data Catalog: Gaia DR2 I/345}}
  (\bibinfo{year}{VizieR, 2018}).

\bibitem{Ting2018}
\bibinfo{author}{{Ting}, Y.-S.}, \bibinfo{author}{{Hawkins}, K.} \&
  \bibinfo{author}{{Rix}, H.-W.}
\newblock \bibinfo{title}{{A Large and Pristine Sample of Standard Candles
  across the Milky Way: 100,000 Red Clump Stars with 3\% Contamination}}.
\newblock \emph{\bibinfo{journal}{Astrophys. J.}}
  \textbf{\bibinfo{volume}{858}}, \bibinfo{pages}{L7} (\bibinfo{year}{2018}).

\bibitem{albareti2017}
\bibinfo{author}{{Albareti}, F.~D.} \emph{et~al.}
\newblock \bibinfo{title}{{The 13th Data Release of the Sloan Digital Sky
  Survey: First Spectroscopic Data from the SDSS-IV Survey Mapping Nearby
  Galaxies at Apache Point Observatory}}.
\newblock \emph{\bibinfo{journal}{Astrophys. J. Suppl.}}
  \textbf{\bibinfo{volume}{233}}, \bibinfo{pages}{25} (\bibinfo{year}{2017}).

\bibitem{charbonnel2007}
\bibinfo{author}{{Charbonnel}, C.} \& \bibinfo{author}{{Zahn}, J.~P.}
\newblock \bibinfo{title}{{Thermohaline mixing: a physical mechanism governing
  the photospheric composition of low-mass giants}}.
\newblock \emph{\bibinfo{journal}{Astron. \& Astrophys.}}
  \textbf{\bibinfo{volume}{467}}, \bibinfo{pages}{L15--L18}
  (\bibinfo{year}{2007}).

\bibitem{Lattanzio2015}
\bibinfo{author}{{Lattanzio}, J.~C.} \emph{et~al.}
\newblock \bibinfo{title}{{On the numerical treatment and dependence of
  thermohaline mixing in red giants}}.
\newblock \emph{\bibinfo{journal}{Mon. Not. R. Astron. Soc.}}
  \textbf{\bibinfo{volume}{446}}, \bibinfo{pages}{2673--2688}
  (\bibinfo{year}{2015}).

\bibitem{lind2009}
\bibinfo{author}{{Lind}, K.}, \bibinfo{author}{{Asplund}, M.} \&
  \bibinfo{author}{{Barklem}, P.~S.}
\newblock \bibinfo{title}{{Departures from LTE for neutral Li in late-type
  stars}}.
\newblock \emph{\bibinfo{journal}{Astron. \& Astrophys.}}
  \textbf{\bibinfo{volume}{503}}, \bibinfo{pages}{541--544}
  (\bibinfo{year}{2009}).

\bibitem{Angelou2012}
\bibinfo{author}{{Angelou}, G.~C.}, \bibinfo{author}{{Stancliffe}, R.~J.},
  \bibinfo{author}{{Church}, R.~P.}, \bibinfo{author}{{Lattanzio}, J.~C.} \&
  \bibinfo{author}{{Smith}, G.~H.}
\newblock \bibinfo{title}{{The Role of Thermohaline Mixing in Intermediate- and
  Low-metallicity Globular Clusters}}.
\newblock \emph{\bibinfo{journal}{Astrophys. J.}}
  \textbf{\bibinfo{volume}{749}}, \bibinfo{pages}{128} (\bibinfo{year}{2012}).

\bibitem{Dorazi2015}
\bibinfo{author}{{D'Orazi}, V.} \emph{et~al.}
\newblock \bibinfo{title}{{Lithium abundances in globular cluster giants: NGC
  1904, NGC 2808, and NGC 362}}.
\newblock \emph{\bibinfo{journal}{Mon. Not. R. Astron. Soc.}}
  \textbf{\bibinfo{volume}{449}}, \bibinfo{pages}{4038--4047}
  (\bibinfo{year}{2015}).

\bibitem{Kirby2016}
\bibinfo{author}{{Kirby}, E.~N.} \emph{et~al.}
\newblock \bibinfo{title}{{Lithium-rich Giants in Globular Clusters}}.
\newblock \emph{\bibinfo{journal}{Astrophys. J.}}
  \textbf{\bibinfo{volume}{819}}, \bibinfo{pages}{135} (\bibinfo{year}{2016}).

\bibitem{paxton2013}
\bibinfo{author}{{Paxton}, B.} \emph{et~al.}
\newblock \bibinfo{title}{{Modules for Experiments in Stellar Astrophysics
  (MESA): Planets, Oscillations, Rotation, and Massive Stars}}.
\newblock \emph{\bibinfo{journal}{Astrophys. J. Suppl.}}
  \textbf{\bibinfo{volume}{208}}, \bibinfo{pages}{4} (\bibinfo{year}{2013}).

\bibitem{AnthonyTwarog2018}
\bibinfo{author}{{Anthony-Twarog}, B.~J.}, \bibinfo{author}{{Lee-Brown},
  D.~B.}, \bibinfo{author}{{Deliyannis}, C.~P.} \& \bibinfo{author}{{Twarog},
  B.~A.}
\newblock \bibinfo{title}{{WIYN Open Cluster Study. LXXVI. Li Evolution Among
  Stars of Low/Intermediate Mass: The Metal-deficient Open Cluster NGC 2506}}.
\newblock \emph{\bibinfo{journal}{Astron. J.}} \textbf{\bibinfo{volume}{155}},
  \bibinfo{pages}{138} (\bibinfo{year}{2018}).

\bibitem{pilachowski1986}
\bibinfo{author}{{Pilachowski}, C.}
\newblock \bibinfo{title}{{The abundance of lithium in old galactic clusters. I
  - NGC 7789}}.
\newblock \emph{\bibinfo{journal}{Astrophys. J.}}
  \textbf{\bibinfo{volume}{300}}, \bibinfo{pages}{289--296}
  (\bibinfo{year}{1986}).

\bibitem{aguileragomez2016}
\bibinfo{author}{{Aguilera-G{\'o}mez}, C.}, \bibinfo{author}{{Chanam{\'e}},
  J.}, \bibinfo{author}{{Pinsonneault}, M.~H.} \& \bibinfo{author}{{Carlberg},
  J.~K.}
\newblock \bibinfo{title}{{On Lithium-rich Red Giants. I. Engulfment of
  Substellar Companions}}.
\newblock \emph{\bibinfo{journal}{Astrophys. J.}}
  \textbf{\bibinfo{volume}{829}}, \bibinfo{pages}{127} (\bibinfo{year}{2016}).

\bibitem{deepak2020}
\bibinfo{author}{{Deepak}}, \bibinfo{author}{{Lambert}, D.~L.} \&
  \bibinfo{author}{{Reddy}, B.~E.}
\newblock \bibinfo{title}{{Abundance analyses of Li-enriched and normal giants
  in the GALAH survey}}.
\newblock \emph{\bibinfo{journal}{Mon. Not. R. Astron. Soc.}}
  \textbf{\bibinfo{volume}{494}}, \bibinfo{pages}{1348--1365}
  (\bibinfo{year}{2020}).

\bibitem{cayrel1988}
\bibinfo{author}{{Cayrel}, R.}
\newblock \bibinfo{title}{{Data Analysis}}.
\newblock In \bibinfo{editor}{{Cayrel de Strobel}, G.} \&
  \bibinfo{editor}{{Spite}, M.} (eds.) {\bibinfo{booktitle}{The Impact of
  Very High S/N Spectroscopy on Stellar Physics}}, \emph{\bibinfo{series}{IAU Symp.}}, \textbf{\bibinfo{volume}{132}}, \bibinfo{pages}{345}
  (\bibinfo{year}{1988}).

\bibitem{sneden1973}
\bibinfo{author}{{Sneden}, C.~A.}
\newblock {\bibinfo{title}{{Carbon and Nitrogen Abundances in Metal-Poor
  Stars.}}}
\newblock Ph.D. thesis, \bibinfo{school}{THE UNIVERSITY OF TEXAS AT AUSTIN.}
  (\bibinfo{year}{1973}).

\bibitem{Gilmore2012}
\bibinfo{author}{{Gilmore}, G.} \emph{et~al.}
\newblock \bibinfo{title}{{The Gaia-ESO Public Spectroscopic Survey}}.
\newblock \emph{\bibinfo{journal}{The Messenger}}
  \textbf{\bibinfo{volume}{147}}, \bibinfo{pages}{25--31}
  (\bibinfo{year}{2012}).

\bibitem{kumar2018}
\bibinfo{author}{{Kumar}, Y.~B.}, \bibinfo{author}{{Reddy}, B.~E.} \&
  \bibinfo{author}{{Zhao}, G.}
\newblock \bibinfo{title}{{Identifying Li-rich giants from low-resolution
  spectroscopic survey}}.
\newblock \emph{\bibinfo{journal}{J. Astrophys. Astron.}}
  \textbf{\bibinfo{volume}{39}}, \bibinfo{pages}{25} (\bibinfo{year}{2018}).

\bibitem{mosser2012}
\bibinfo{author}{{Mosser}, B.} \emph{et~al.}
\newblock \bibinfo{title}{{Probing the core structure and evolution of red
  giants using gravity-dominated mixed modes observed with Kepler}}.
\newblock \emph{\bibinfo{journal}{Astron. \& Astrophys.}}
  \textbf{\bibinfo{volume}{540}}, \bibinfo{pages}{A143} (\bibinfo{year}{2012}).

\bibitem{Takeda2017}
\bibinfo{author}{{Takeda}, Y.} \& \bibinfo{author}{{Tajitsu}, A.}
\newblock \bibinfo{title}{{On the observational characteristics of
  lithium-enhanced giant stars in comparison with normal red giants}}.
\newblock \emph{\bibinfo{journal}{Publ. Astron. Soc. Jpn.}}
  \textbf{\bibinfo{volume}{69}}, \bibinfo{pages}{74} (\bibinfo{year}{2017}).

\bibitem{fulbright00}
\bibinfo{author}{{Fulbright}, J.~P.}
\newblock \bibinfo{title}{{Abundances and Kinematics of Field Halo and Disk
  Stars. I. Observational Data and Abundance Analysis}}.
\newblock \emph{\bibinfo{journal}{Astron. J.}} \textbf{\bibinfo{volume}{120}},
  \bibinfo{pages}{1841--1852} (\bibinfo{year}{2000}).

\bibitem{lambert91}
\bibinfo{author}{{Lambert}, D.~L.}, \bibinfo{author}{{Heath}, J.~E.} \&
  \bibinfo{author}{{Edvardsson}, B.}
\newblock \bibinfo{title}{{Lithium abundances for 81 F dwarfs.}}
\newblock \emph{\bibinfo{journal}{Mon. Not. R. Astron. Soc.}}
  \textbf{\bibinfo{volume}{253}}, \bibinfo{pages}{610--610}
  (\bibinfo{year}{1991}).

\bibitem{girardi2016}
\bibinfo{author}{{Girardi}, L.}
\newblock \bibinfo{title}{{Red Clump Stars}}.
\newblock \emph{\bibinfo{journal}{Ann. Rev. Astron. Astrophys.}}
  \textbf{\bibinfo{volume}{54}}, \bibinfo{pages}{95--133}
  (\bibinfo{year}{2016}).

\end{thebibliography}


\begin{addendum}
\item 
This work made use of the GALAH survey, which includes data acquired through the Australian Astronomical Observatory. It also made use of the astronomical data analysis software TOPCAT, and the NASA Astrophysics Data Service (ADS). Y.B.K. and G.Z. acknowledge the support of the National Natural Science Foundation of China through grant numbers 11988101, 11850410437, 11890694 and the National Key R$\&$D Program of China grant number 2019YFA0405502. B.E.R. thanks NAOC, Beijing, for support through the CAS PIFI grant number 2019VMA0009. S.W.C. acknowledges federal funding from the Australian Research Council through a Future Fellowship (FT160100046) and Discovery Project (DP190102431). Y.-S.T. acknowledges support from the NASA Hubble Fellowship grant HST-HF2-51425.001 awarded by the Space Telescope Science Institute. We also thank L. Spina and C. Doherty for discussions.

\item[Author Contributions]
Y.B.K , B.E.R, and G.Z proposed and designed this study. Y.B.K analysed the data, and prepared the manuscript. B.E.R assisted in data analysis and manuscript preparation. S.W.C calculated the stellar models, assisted in the data analysis, and prepared the manuscript. S.M. assisted in data analysis and preparing the figures. G.Z. assisted in manuscript preparation. Y.S.T. made the asteroseismic inference catalogues, and assisted in manuscript preparation. 

\item[Author Information] Reprints and permissions information is available at www.nature.com/reprints. The authors declare that they have no competing financial interests. The data that support the plots within this paper and other findings of this study are available from the corresponding author upon reasonable request. Correspondence and requests for materials should be addressed to Y.B.K. or G.Z.
\end{addendum}

\newpage

\begin{figure}
  \begin{center}
     \includegraphics[width=0.9\textwidth]{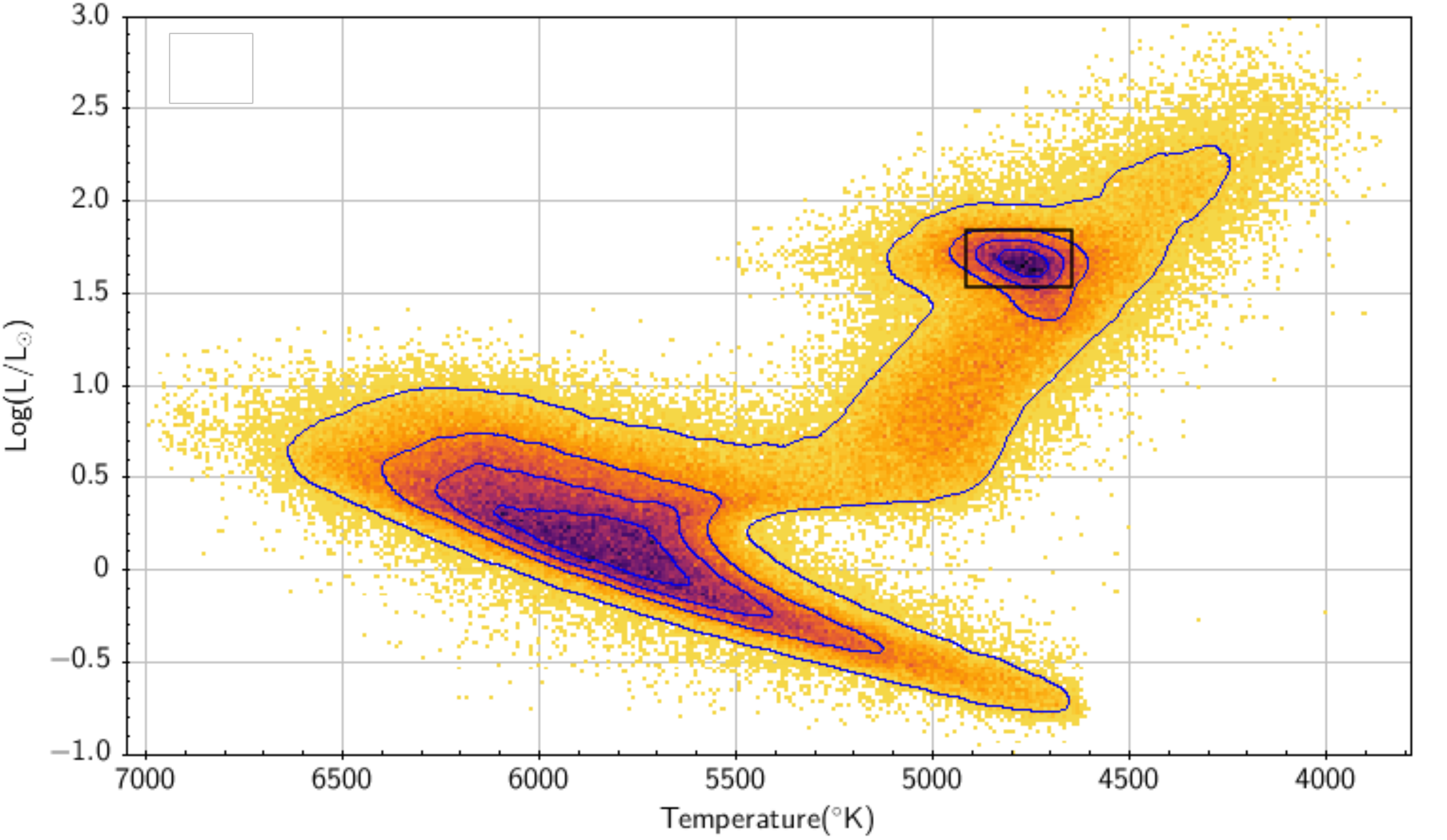}
\caption{Hertzsprung-Russell diagram of the Galah DR2 sample used to locate the red clump stars. The small rectangle marks our red clump box region. It coincides with the high number-density contours around the red clump, avoiding the RGB bump overdensity just below. The colour scale is linear and represents the number density of stars, a darker colour indicating a higher density. The contours reflect the same density scale. The RC box covers a narrow luminosity range log(L/L$_{\odot}) = 1.55 \rightarrow 1.85$ and surface temperature range T$_{\rm{eff}} = 4650 \rightarrow 4900$~K. The RC box limits match well with theoretical predictions of the core He-burning phase for low-mass stars\cite{girardi2016}. Only stars with high quality surface temperature determinations (with Galah flag$_{\rm{cannon}} = 0$) and with masses $< 2.0$~M$_{\odot}$ are used, giving a total sample of 225,580 low-mass stars in the diagram. Luminosities are from the GAIA DR2 catalogue\cite{GAIADR2Cat}, whilst the surface temperatures are from the Galah DR2 catalogue\cite{buder2018}.}
     \label{fig:fig1}
     \end{center}
 \end{figure}
 \begin{figure}
     \centering
     \includegraphics[width=0.9\textwidth]{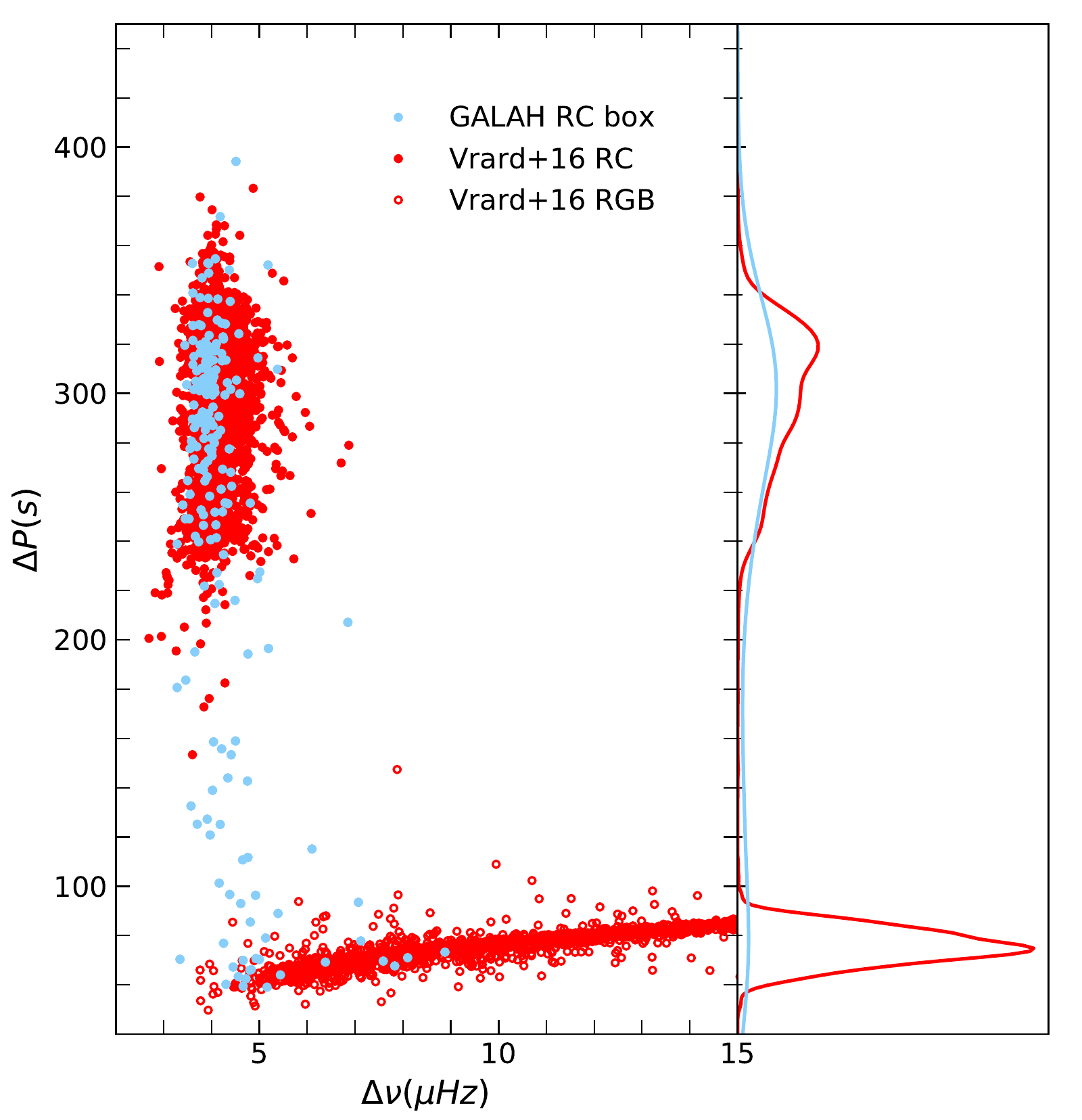}
     \caption{Asteroseismic properties of giant stars. Blue symbols show the overlap of the Galah sample with an asteroseismic parameter catalogue\cite{Ting2018} (inferred parameters; based on LAMOST and APOGEE spectra). Red symbols are Kepler RGB and RC stars with direct asteroseismic measurements (Vrard+16 RC and Vrard+16 RGB, respectively)\cite{vrard2016}. The contamination of RGB stars among blue symbols is $10\%$. The histograms at the right show the distributions of $\Delta P$ in each dataset, with corresponding colours.}
     \label{fig:fig2}
 \end{figure}
 
 \begin{figure}
     \centering
     \includegraphics[width=0.9\textwidth]{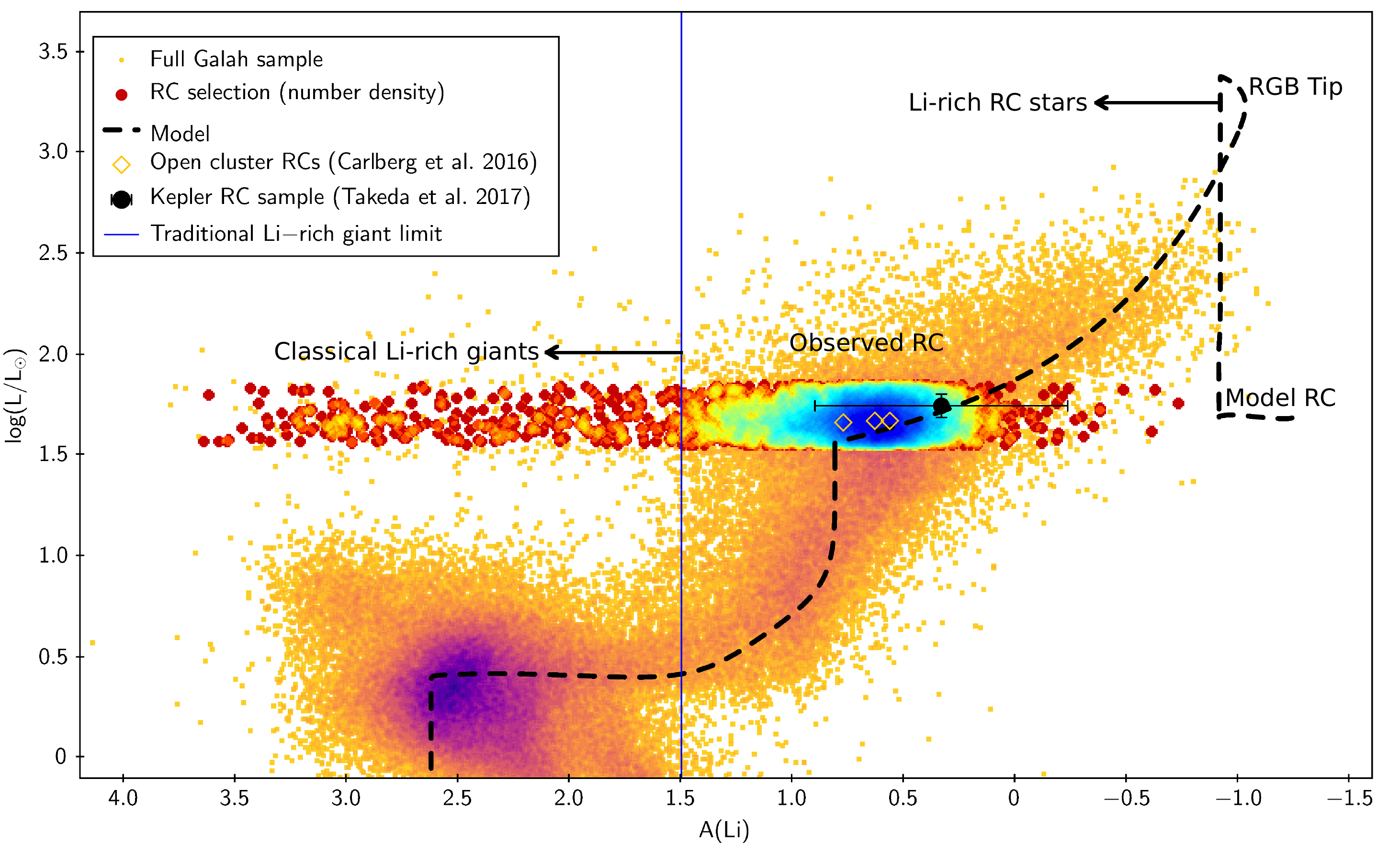}
     \caption{Our Galah stellar sample in the lithium-luminosity plane. The sample is based on Figure~1, but is limited to stars with reliable Li determinations (Galah flag$_{[\rm{Li/Fe}]} \leq 3$, and errors of A(Li) $< 0.3$ dex). The sample has 107,776 stars ('Full Galah sample'). The colour scale for this sample represents the number density of stars, a darker colour indicating a higher density. This diagram shows the change in abundance of lithium as the stars evolve. It can be seen that stars at the RGB tip (top right) have the lowest Li abundances, since stars generally deplete Li during their evolution. Stars in all other phases of evolution have A(Li) greater than the RGB tip. The RC stars, which were identified in the Figure~1 RC box, spread out in A(Li), forming the band of stars at the RC luminosity. This RC band is highlighted by the blue-to-red number density colour scale, where blue represents highest density. It can be seen that the number density of RC stars, which is peaked at A(Li)~$=+0.71$~dex ($\sigma = 0.39$~dex; `Observed RC'). This contrasts strongly with the model prediction (black dashed line) for the RC location, which begins at A(Li)~$=-0.9$~dex, and then reduces further (`Model RC'). It is clear that the model does not match the observed RC distribution, with a difference of $\Delta$A(Li)~$=1.6$~dex (factor of 40). The solid vertical line represents the canonical first dredge-up Li upper limit\cite{iben1967a} (A(Li)~$> +1.5$~dex) used for defining the classical Li-rich giants. These stars represent only the tail of the RC distribution, lying outside $2\sigma$ of the mainstream RC lithium distribution. 
     Shown for comparison are average Li abundances for RC star samples in three open clusters (NGC2204, NGC2506, and Collinder 110), taken from the literature\cite{carlberg2016}. The typical 1$\sigma$ range of A(Li) in these samples is $\sim 0.2$~dex. These OC stars have masses within our sample's range (M~$\lesssim 2.0$~M$_{\odot}$), and have metallicities from solar to $-0.41$~dex. Also shown for comparison is the average Li abundance from a sample of RC stars taken from the Kepler space telescope's original field of view\cite{Takeda2017}. The horizontal error bars show the 1$\sigma$ A(Li) range of the sample. These independent samples support our finding. See Extended Fig.~E3 for the individual star abundances of the literature samples.
     The stellar model was computed with the MESA stellar code\cite{paxton2013}, with parameters: initial mass = 1.0~M$_{\odot}$, [Fe/H]~$=0.0$, and thermohaline mixing efficiency\cite{paxton2013} $\alpha_{thm}= 100$.
     }
     \label{fig:fig3}
\end{figure}

 \begin{figure}
 \begin{center}
  \includegraphics[width=1.0\textwidth]{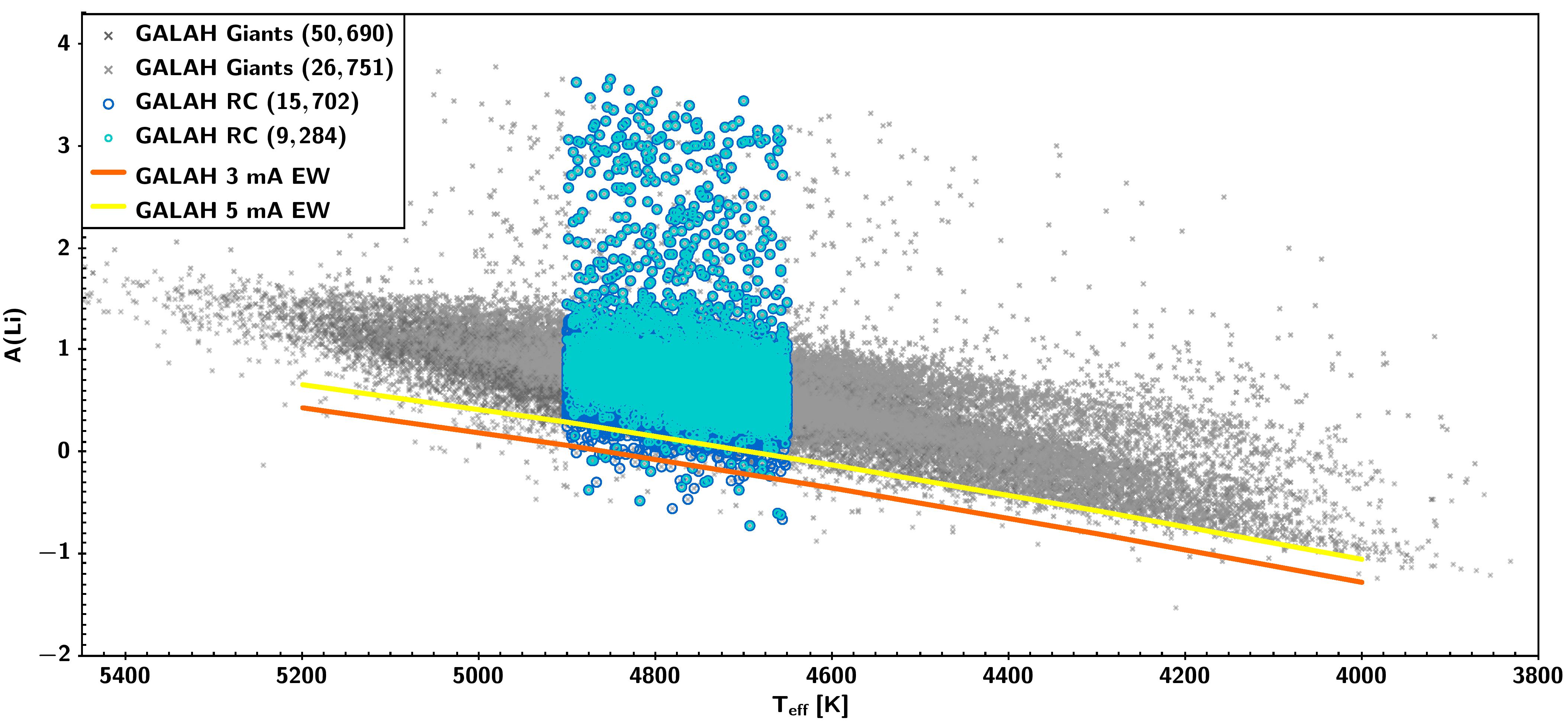} 
  \captionsetup{labelformat=empty}
  \caption{{\bf Figure~E1.} Li vs T$_{\rm{eff}}$ of giants ($\log(g)<3.1$). The Galah sample is shown without reduction by lithium quality flag (dark grey crosses, 50,690 stars), and with Li quality flag as used in manuscript (light grey crosses, 26,751 stars). The respective red clump (RC) sub-samples are highlighted by blue and cyan symbols. Our Li detection curves, computed using spectral synthesis, are shown by the solid lines (see text for details). It can be seen that the vast majority of stars, in all samples, are above the detection limits. See also Extended Figure~E2, which provides a cross-section histogram of this diagram in the red clump T$_{\rm{eff}}$ range.}
  \label{figE1}
 \end{center}
 \end{figure}

  \begin{figure}
  \centering
  \includegraphics[width=1.0\columnwidth]{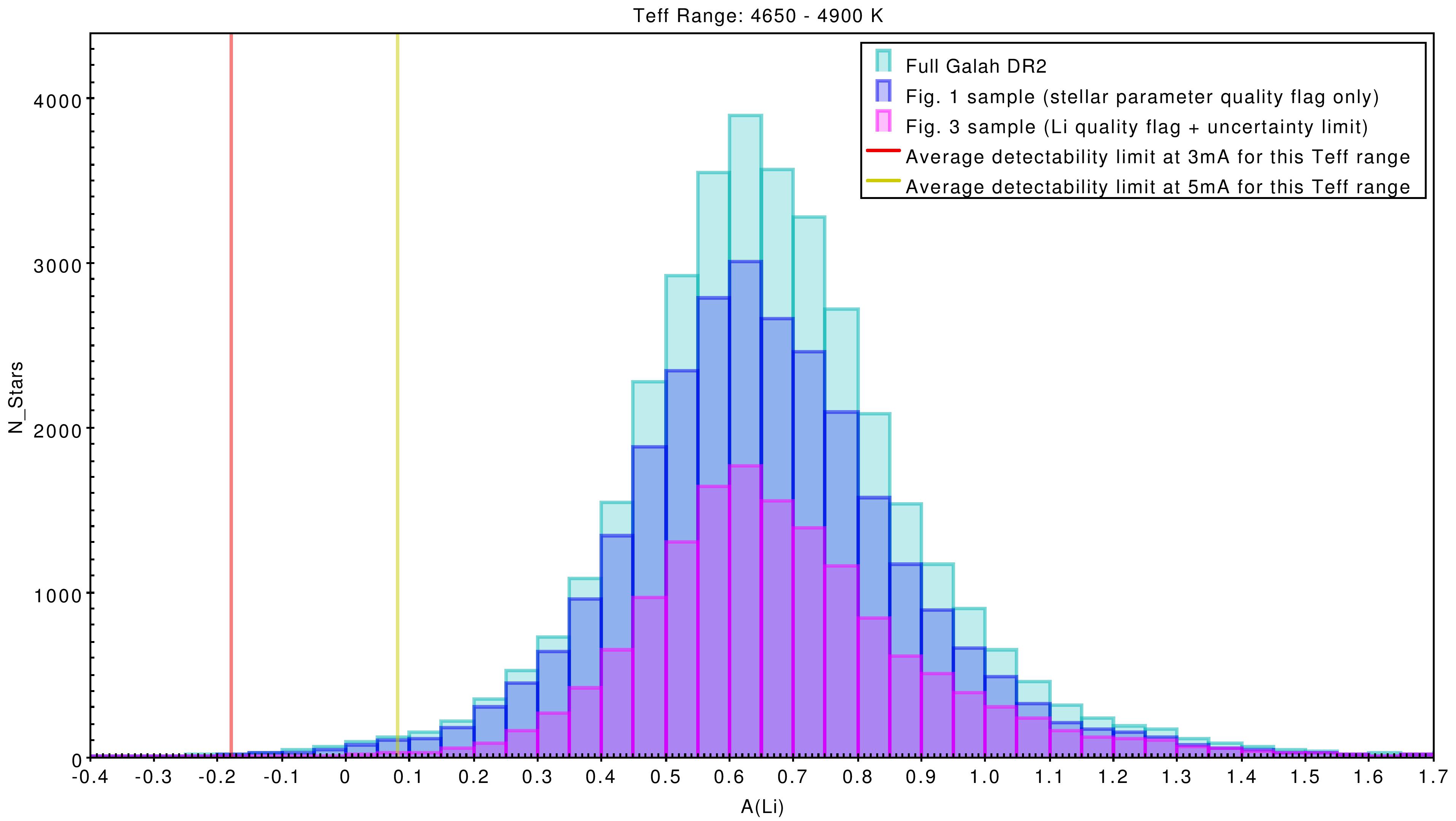} 
  \captionsetup{labelformat=empty}
  \caption{{\bf Figure~E2.} Histograms showing number of giant stars per A(Li) bin. The stellar sample has restricted  T$_{\rm{eff}} = 4650 - 4900$~K, which matches our RC T$_{\rm{eff}}$ range. We have not restricted the samples in any other way, that is, they contain all stars in this range, regardless of evolutionary state (RC or RGB). The largest sample is the full Galah DR2 (35,800 stars), the next largest is the sample for manuscript Fig.~1 (27,847 stars), and the smallest is from our manuscript Fig.~3 sample (15,469 stars). The vertical lines show our computed detectability limits (see Extended Fig.~1, and text for details). It can be seen that the bulk of the distribution is well above the detection limit, and that the location of the peak is unchanged through each sub-sampling.}
  \label{figE2}
  \end{figure}
  
  \begin{figure}
  \centering
  \includegraphics[width=1.0\columnwidth]{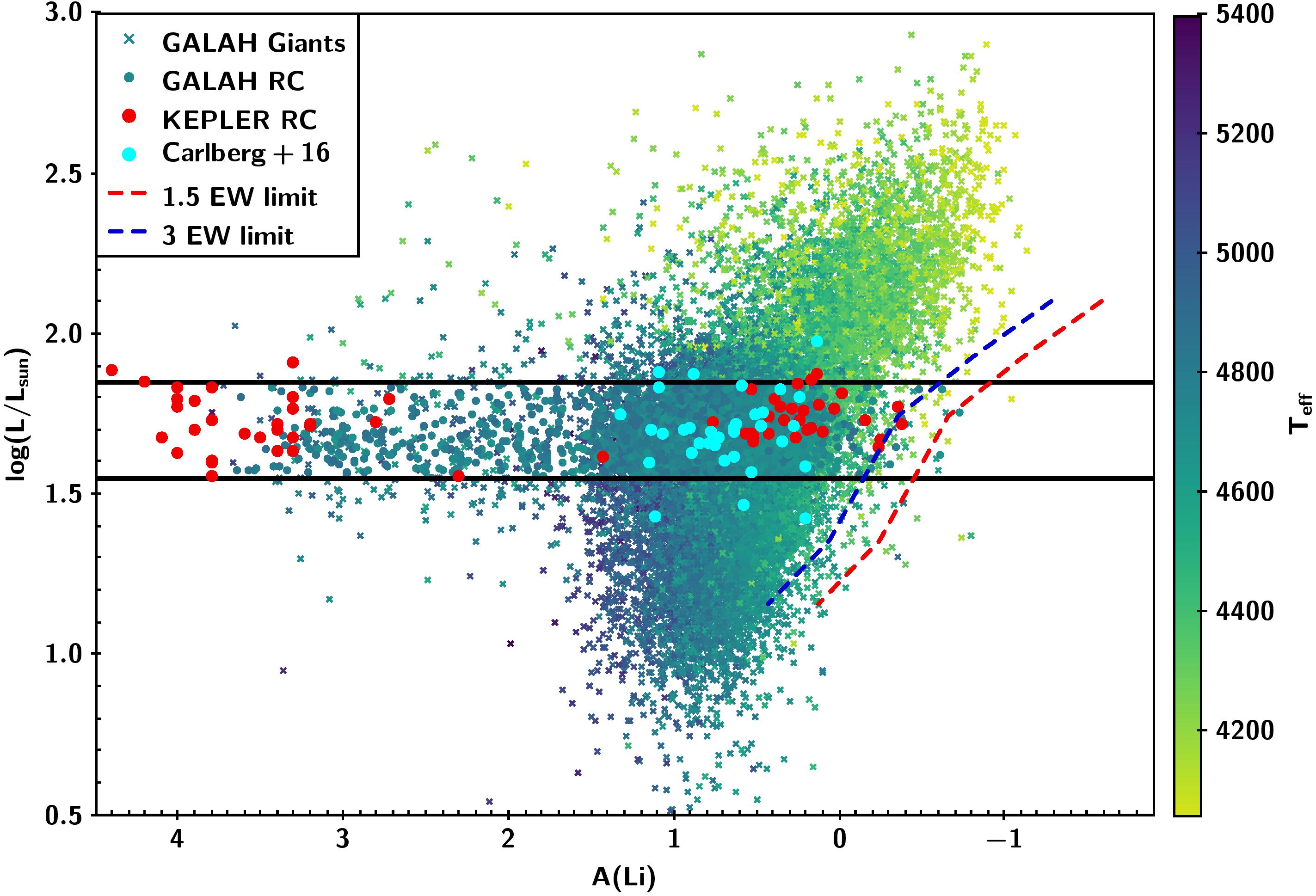} 
  \captionsetup{labelformat=empty}
  \caption{{\bf Figure~E3.} Li vs Luminosity of giants. Color gradient shows the temperature distribution. Red circles are Kepler RC stars, with the lower-Li sample\cite{mosser2012,Takeda2017} and super-Li-rich sample\cite{singh2019b} taken from separate studies. Light blue circles are individual RC stars from open clusters\cite{carlberg2016}. Li detection limits computed for GALAH and the lower-Li Kepler RC sample are shown in blue and red dashed lines, respectively. It can be seen that the peaks of these literature distributions are much higher than the RGB tip value of $\sim -1.0$~dex, as found in our Galah sample. The peaks are also well above the respective detectability limits.}
  \label{figE3}
  \end{figure}
  
  \begin{figure}
   \centering
   \includegraphics[width=1.0\columnwidth]{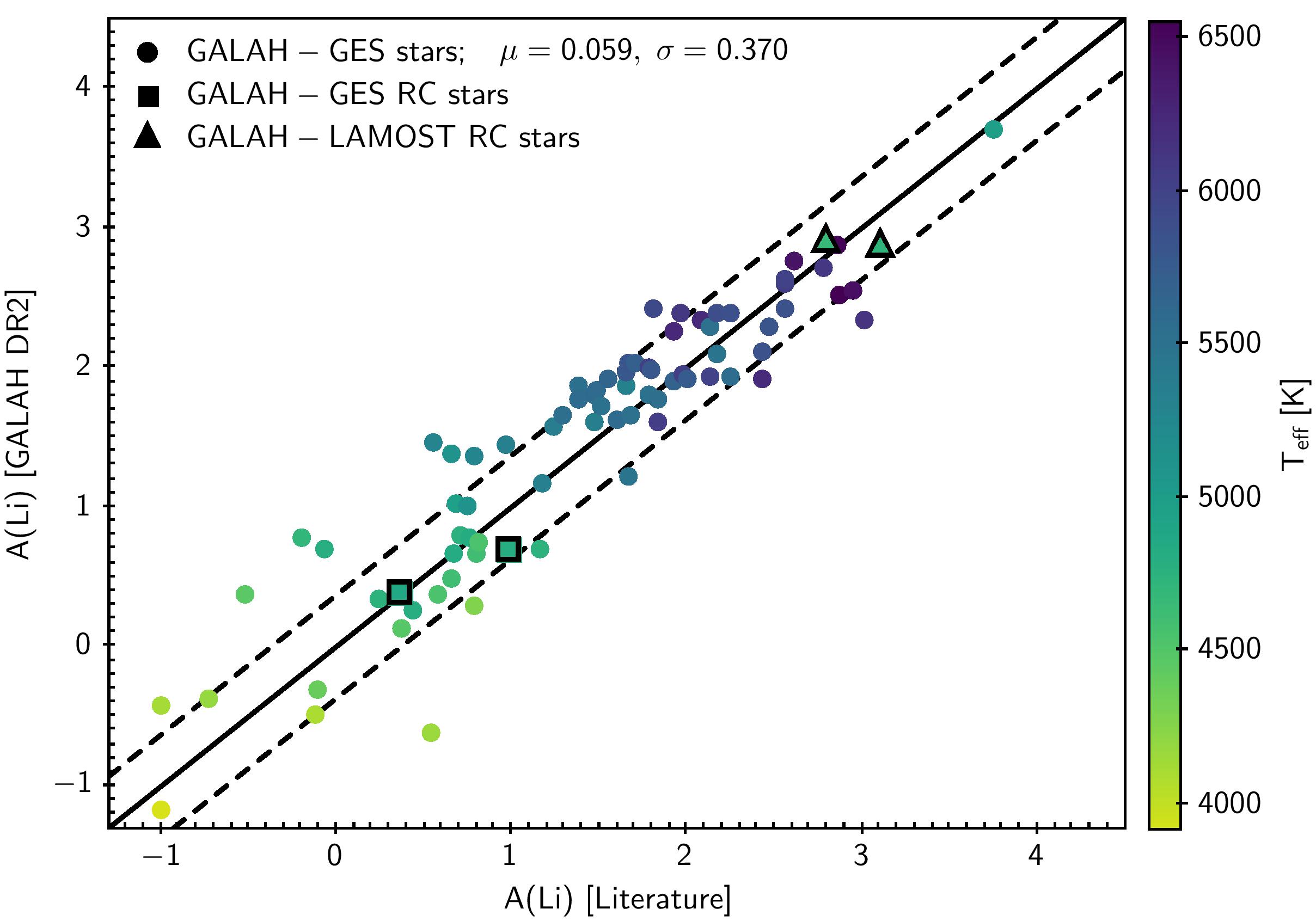}
   \captionsetup{labelformat=empty}
  \caption{{\bf Figure~E4.} Comparison of lithium abundances between Galah DR2 and Gaia-ESO survey\cite{Gilmore2012} DR3 (`GES'). A total of 78 stars were found in common (circles). A very slight offset between Gaia-ESO and Galah is found: $+0.059$~dex ($\sigma=0.37$). The common sample includes two red clump stars,
  indicated by squares. These RC stars straddle the peak of the Galah RC distribution at A(Li)~$=0.71$~dex. 
  Also shown are two RC giants found in a cross-match between Galah DR2 and the LAMOST survey\cite{cui2012} (triangles). As LAMOST is a low-resolution survey, Li is only detectable in very Li-rich stars. The colour gradient shows the temperature distribution.}
   \label{figE4}
  \end{figure}

\end{document}